\documentstyle[prb,preprint,aps]{revtex}
\begin{document}
\draft
\preprint{\footnotesize{ Applied Physics Report 94-34}}
\title{
Influence of Long-Range Coulomb Forces and Dissipation on the Persistent
Current of a Wigner Crystal-Ring
}
\author{I.~V.~Krive$^{(1,2)}$, P.~Sandstr\"{o}m$^{(1)}$,
R.~I.~Shekhter$^{(1)}$,  and M. Jonson$^{(1)}$}
\address{
$^{(1)}$Department of Applied Physics, Chalmers University of
Technology and G\"{o}teborg University, S-412 96 G\"{o}teborg,
Sweden }
\address{
$^{(2)}$B. I. Verkin Institute for Low Temperature Physics and Engineering,
47 Lenin Avenue, 310164 Kharkov, Ukraine
}
\maketitle

\begin{abstract}
The influence of long-range Coulomb forces on the persistent current
of a pinned Wigner crystal-ring is considered. A simple method
is proposed for how to take into account Coulomb effects for the macroscopic
quantum tunneling of the Wigner crystal. It is shown that unscreened
Coulomb interactions increase the stiffness of the Wigner lattice. This leads
to additional suppression of the amplitude of Aharonov-Bohm oscillations, but
makes the predicted anomalous temperature dependence of the persistent
current more pronounced. The impact of dissipation on the persistent current
of a Wigner crystal-ring is also studied.
\end{abstract}
\newpage
In condensed matter the Aharonov-Bohm effect manifests itself
through magnetic oscillations of thermodynamic and kinetic properties of
nonsimply connected mesoscopic systems. It is a well-known fact that the period
of mesoscopic Aharonov-Bohm oscillations is equal to the flux quantum $\Phi_{0}
= hc/e$ ($e$ is the electron charge) and that the oscillation amplitude depends
on the properties (conductivity) of the system.

Until recently the main theoretical and experimental efforts were directed
towards investigating conductance
oscillations in small-sized metallic wires and rings (see e.g. the review [1]).
Succesful measurements of the flux-induced magnetic moment (persistent
current) of isolated single metallic- [2] and semiconductor [3] rings have made
the study of thermodynamic properties of mesoscopic quantum rings a
central problem of considerable current interest. Whereas for metallic rings
the problem at hand amounts to properly  taking into account the effect of
impurities on the persistent current of non- (or weakly) interacting electrons,
the main goal in the case of semiconductor rings --- formed in a laterally
confined high mobility two-dimensional electron gas (2DEG) ---  is to elucidate
the role of electron-electron correlations in the Aharonov-Bohm oscillations.

We recall [4] that in a perfect (impurity free) ring even strong
Coulomb correlations do not influence the persistent current, which is the same
as if calculated in the naive model of free electrons [5, 6]. This result can
be
proved as a theorem for Gallilean invariant systems (at zero temperature) [7].
In
lattice models a drastic change in the magnitude of the persistent current
comes
about only in the vicinity of a metal-insulator phase transition induced by
electron-electron correlations. In the latter case the oscillation amplitude
becomes exponentially small [8,9,10], which is a characteristic feature of the
Aharonov-Bohm effect in dielectrics [11, 12].

It is reasonable to study the quantum dynamics of electrons in one-dimensional
(1D) channels  --- formed in a 2DEG by a smooth electrostatic potential --- in
the framework of continuum models. In this case Coulomb correlations  favor the
formation of an ordered structure at low
density and temperatures. Such a Wigner crystal can exist in rings of
large enough (mesoscopic) sizes [13]. A simple model of a 1D Wigner crystal has
been suggested in Ref.~[14] and has proven to be suitable for studying charge
transport in strongly correlated electron systems. This model was used in
Ref.~[15] for an evaluation of the persistent current in a ring-shaped Wigner
crystal. It was demonstrated that in a perfect ring the persistent current
caused by the sliding of a Wigner crystal is indistinguishable from the current
carried by free electrons. The situation is drastically changed in the presence
of even a small potential barrier. The barrier pins the crystal and charge
transport is possible only due to the processes of thermal or quantum
depinning. At low temperatures the latter is dominant and the current becomes
of the tunneling type.

For strong pinning both the zero temperature value and the temperature
dependence of the oscillation amplitude are significantly changed.
An increase in temperature favours depinning of the crystal; hence at low
temperatures $T \alt T_{s} \equiv \hbar s/L$ the current increases with
temperature ($s$ is the sound velocity in the Wigner lattice, $L$ is the
circumference of the ring). However, in the high temperature region ($T>T_{s}$)
the persistent current is strongly (exponentially) suppressed due a
temperature enhancement of destructive interference effects (loss of quantum
coherence). Thus we get an anomalous temperature behaviour of the Aharonov-Bohm
oscillations in a pinned Wigner crystal-ring that contradicts the prediction
of Fermi-liquid theory [6] and is specific for strongly correlated
electrons.

The above mentioned results were obtained in Ref.~[15] for a simple model of a
Wigner crystal without long-range Coulomb interactions and without effects of
(possible) dissipative motion of the Wigner lattice. The purpose of the present
paper is to estimate the influence of unscreened Coulomb forces and dissipation
on the persistent current of a pinned Wigner crystal-ring.

A 1D Wigner crystal pinned by a potential $ V_{p}$ --- local on the scale of
the
system size $L$ --- is described by the Lagrangian
\begin{equation}
 {\cal L} =\frac{ma}{8\pi^{2}}\left\{\dot{\varphi}^{2} - s^{2}
(\varphi^{\prime})^{2}\right\} - V_{p}\delta(x)(1-\cos\varphi) +
\frac{\hbar}{L}
\frac{\Phi}{\Phi_{0}} \dot{\varphi}
\end{equation}
Here $a$ is the Wigner lattice spacing ($a^{-1} \equiv n, \;n $ being the
electron density), $m$ is the electron mass, $s$ is the sound velocity (the
velocity of charged excitations - plasmons in the Wigner crystal),
$\varphi\equiv 2\pi u(x,t)/a$ stands for the dimensionless displacement field.
The last term in Eq.(1) describes the Aharonov-Bohm interaction of the Wigner
crystal with the magnetic flux $\Phi$. It is worth  noticing that the
Lagrangian (1) is identical with the bosonized form of the Luttinger model (for
spinless electrons) if one identifies the parameter $\eta_{\rho}$, which
specifies the exponents of different correlation functions in the Luttinger
model (see e.g. Ref.~[16]), with the quantity $\alpha\equiv\pi\hbar /msa$ of
our
model. The parameter $\alpha$ determines the ``strength" of quantum
fluctuations
in the system and in what follows it is assumed to be small $\alpha\ll 1 $
(corresponding to a stiff Wigner crystal).

The long-range part of the Coulomb interaction is decribed by a standard
expression, which in terms of the displacement field $\varphi$ looks as follows
[17] (see also [18])
\begin{equation}
E_{c}=\frac{ e^{2} }{ 8\pi^{2} }\int\int dxdx^{\prime}
\frac {\partial_{x}\varphi(x)\partial_{x^{\prime}}\varphi (x^{\prime}) }
{ \sqrt{(x-x^{\prime})^{2} + d^{2}} } .
\end{equation}
Here $ d $ is the length associated with short wavelength ``screening" (in
fact the width of the one-mode channel). We will take it to be of the order of
the Wigner crystal spacing $ d\alt a $. In this case plasmons in the Wigner
crystal ($ka \ll 1$) experience the influence of unscreened Coulomb forces and
their dynamics are significantly changed.

In a constant-potential region the impact of Coulomb interactions, Eq.(2), on
the  dynamics of plasmons can be accounted for exactly (see e.g. Ref.~[17])
\begin{equation}
\omega(k)=sk \left\{ 1+\frac{ 2e^{2} }{ mas^{2} }K_{0}(ka)\right\}^{1/2}
\stackrel{\rm (ka \ll 1) }{\approx} s_{0}k\sqrt{ \ln(1/ka) }  ,
\end{equation}
where $ K_{0}(x)$ is the MacDonald function, $s_{0}^{2}\equiv 2e^{2}/ma$.

Our prime interest is in the question of how the Coulomb interaction, Eq.(2),
influences the tunneling dynamics of a pinned Wigner crystal-ring. We will
assume
that the circumference of the ring is large enough, $ L \gg a$, so that the
Coulomb interaction becomes essential on scales for which one can neglect the
effects of curvature (the distinction between ring-like and linear geometry).

It is physically evident that the main effect of the long-range Coulomb forces
is connected with the change in the dispersion relation for plasmons,
Eq.(3). How will this modification influence the tunneling dynamics of a Wigner
crystal-ring? It is obvious that Coulomb interactions affect the tunneling
of the Wigner crystal through the barrier if it is accompanied by
a distortion of the crystal so that $\varphi^{\prime}\neq 0$. As was shown in
Ref.~[19], the tunneling process is inhomogeneous in the strong
pinning regime (in our notations  $\alpha_{0}V_{p}\gg T_{s}$ , where
$T_{s}\equiv \hbar s_{0}/L,\; \alpha_{0} \equiv \pi\hbar/mas_{0}\simeq
\sqrt{a_{B}/a},\; a_{B}\equiv\hbar^{2}/e^{2}m $ being the Bohr radius).
In this case the process of depinning the Wigner crystal can be
divided into two different stages. In the first stage, the field
$\varphi$ rapidly changes its value, $\Delta\varphi=\pm 2\pi$, in the region
$l_{0}\simeq ams_{0}^{2}/V_{p}\ll L$ that contains the barrier. The second
stage describes a slow relaxation of the elastic deformation that is a
result of the first stage; the relaxation is associated with propagation of
plasmons. It is this second stage that dominates the action in the
strong-pinning limit and in our case it is modified by the long-range forces.

The instanton describing the tunneling relaxation is the solution of a free
equation of motion (in imaginary time $\tau$) of the form
\begin{equation}
\ddot{\varphi}+s^{2}\varphi^{\prime\prime}+
\frac{e^{2} }{ma}\partial_{x}\left\{ \int_{-L/2}^{L/2}dy
\frac{\partial_{y}\varphi(y,\tau) }{\sqrt{ (x-y)^{2}+a^{2} }}\right\}=0
\end{equation}
For a screened Coulomb interaction the desired solution is of the form [19]
$\varphi=\pm 2\arctan(s\tau/|x|)$. It is easy to show that for $ x\sim  s\tau
\gg a $ the last term in Eq.(4) can be represented in a local form and the
modified Larkin-Lee instanton, $\varphi_{L-L}$, obeys the equation 
\begin{equation}
\ddot{\varphi}_{L-L}+s_{0}^{2}ln\left(\frac{|x|}{a}\right)
\varphi^{\prime\prime}_{L-L}=0,\; \;\;|x| \gg a  .
\end{equation}
(Recall that we placed the barrier at the point $x_{0}=0$). Thus our problem is
reduced to the one solved by Larkin and Lee [19], but now with an $x$-dependent
velocity $s(x)=s_{0}\sqrt{\ln(|x|/a)}$. Since $s(x)$ is a smooth function,
$|s^{\prime}(x)/s(x)|\ll 1$, we readily get the desired instanton solution
\begin{equation}
\varphi_{L-L}(x,\tau)=\pm 2 \arctan
\left(\sqrt{\ln\left(\frac{|x|}{a}\right)} \frac{s_{0}\tau}{|x|} \right)
\end{equation}
For the tunneling action corresponding to the trajectory (6) we find (see
also [14])
\begin{equation}
A_{t}=\frac{\hbar}{\alpha_{0}}\frac{2}{3}
\left\{\ln^{3/2}\left(\frac{L}{a}\right) -
\ln^{3/2}\left(\frac{l_{0}}{a}\right)\right\} .
\end{equation}
The criterion of strong pinning implies that $V_{p} \gg e^{2}/L$. It is
reasonable to assume that the renormalized barrier height --- renormalized by
short wavelength fluctuations --- obeys the inequality $V_{p}<e^{2}/a$. Then
the
persistent current at zero temperature takes the form
\begin{equation}
I_{WC}(T=0) \sim (-1)^{N} \frac{eT_{s}}{\hbar}
\left(\frac{e^{2} }{ LV_{p}}\right)^
\frac{\sqrt{\ln N} }{\alpha_{0}} \sin\left(2\pi \frac{\Phi}{\Phi_{0}}\right) ,
\end{equation}
where nonessential prefactors have been omitted and $N$ is the total number of
electrons in the ring. The appearance of the
additional factor $ \sqrt{\ln N}>1 $ in the exponent in Eq.(8), can easily be
understood from the scaling procedure
\begin{equation}
s\rightarrow s_{0}\ln^{1/2}\left(\frac{L}{a}\right),
\; \alpha^{-1}\sim mas\rightarrow  \alpha_{0}^{-1} \ln^{1/2}(N) .
\end{equation}

Now we consider the temperature behaviour of the persistent current. For a
screened Coulomb interaction this problem has been studied in Ref.~[15]. The
 trick described  above (the replacement $ s\rightarrow s(x)=s_{0}
\sqrt{\ln(|x|/a)}$ ) permits us to use use the results of Ref.~[15] and readily
get the solution. At finite temperature the extremal trajectory is described
by the function [15]
\begin{equation}
\varphi_{\beta}(x,\tau) =\pm 2 \arctan
\left\{\coth\left[\frac{\pi |x|}{\hbar \beta s(x)}\right]
 \tan\left[\frac{\pi}{\hbar\beta}(\tau-\frac{\hbar\beta}{2})\right]\right\} ,
\end{equation}
where $\beta\equiv T^{-1}$ is the inverse temperature. The tunneling action
corresponding to the  periodic instanton (10) is
\begin{equation}
A_{t}(\tau)=\frac{\hbar}{\alpha_{0}} \int_{ \pi\frac{T}{T_i} }
^{ \frac{\pi}{2}\frac{T}{T_{s}} }dx
\left\{\coth\left[\frac{x}{\sqrt{\ln(\frac{T_{c}}{T}x)}}\right] +
        \tanh\left[\frac{x} {\sqrt{\ln( \frac{T_{c}}{T}x)} }\right] -
1\right\} ,
\end{equation}
where $T_{i}\equiv \hbar s_{0}/l_{0},\; T_{c}\equiv \hbar s_{0}/a$. As a
function of temperature, the action (11) attains its minimum at
\begin{equation}
T_{M} \simeq 0.5T_{s}\sqrt{\ln N}, \;\;
A_{t}(T_{M}) \simeq \frac{\sqrt{\ln N}}{\alpha_{0}} ,
\end{equation}
which leads to the anomalous temperature dependence of the amplitude of the
Aharonov-Bohm oscillations (see Fig.1).

Hence, the unscreened Coulomb interaction results in additional suppression of
the persistent current of a strongly pinned Wigner crystal-ring. The peak in
the current plotted as a function of temperature therefore becomes
more pronounced. These results are consistent with the fact that long-range
Coulomb interactions, by suppressing the quantum fluctuations of a Wigner
crystal, transform it into an almost classical system $\alpha_{0}^{-1}
\sqrt{\ln
N} \rightarrow \infty$ as $N \rightarrow \infty$.

Now we proceed to the next question. How does dissipation influence the
persistent current of a Wigner crystal-ring? The first problem is to introduce
dissipation in our model Lagrangian, Eq.(1). Here we will use a
phenomenological
approach and consider the simplest model of dissipative motion, namely linear
friction. In other words we will assume that due to the presumably weak
coupling
of electrons in the ring with the reservoir, the motion of a Wigner crystal
becomes dissipative and can be described in terms of friction in a real time
formalism. In such an approach it is reasonable to consider the friction as
homogenous and local. Then for homogenous translations of the Wigner lattice,
the dissipative equation of motion in real time looks as follows
\begin{equation}
m\ddot{\varphi} + \eta\dot{\varphi} = 0.
\end{equation}
Here $m$ is
the electron mass and $\eta $ is the friction coefficient. This equation, in
essence, determines the normalization of the phenomenological parameter $\eta$.

To include friction in the tunneling dynamics, where imaginary time is more
appropriate, we will follow the standard approach by Caldeira and Legget [20].
According to them linear friction in imaginary time, $\tau$, can be described
by
the nonlocal action
\begin{equation}
A_{C-L} \sim \eta \int_{0}^{\beta}\int_{0}^{\beta} d\tau d\tau^{\prime}
\frac{[\varphi(\tau)-\varphi(\tau^{\prime})]^{2}}
{\frac{\beta^{2}}{\pi^{2}} \sin^{2}[\frac{\pi}{\beta} (\tau-\tau^{\prime})] } .
\end{equation}
Notice, however, that in our model the trajectories in imaginary time which
differ from each other by $2 \pi n$ ($n$ is an integer) are
physically indistinguishable. Therefore the action (14) should be slightly
modified in order to account for the $2\pi$-symmetry of the model, Eq.(1). We
propose a dissipative action of the form
\begin{equation}
A_{d}=\frac{a \eta}{4 \pi^{3}} \int_{-L/2}^{L/2}dx \int_{0}^{\beta}
\int_{0}^{\beta} d\tau d\tau^{\prime} \frac{
\sin^{2}\left[\frac{\varphi(x,\tau)-\varphi(x,\tau^{\prime})}{2}\right] }
{\frac{ \beta^{2} }{\pi^{2}} \sin^{2}[\frac{\pi}{\beta} (\tau-\tau^{\prime})]
},
\end{equation}
which leads to the desired form (13) of the equation of motion  in
real time. It furthermore respects the $2\pi$-symmetry of the quantum dynamics
of the system in question. It is worth noting that the dissipative action
(15) is analogous to the one used repeatedly in superconductivity for
describing the dissipative dynamics of Josephson junctions (see e.g. the review
[21]).

For the nonlinear and nonlocal action , Eqs.~(1) and (14), we failed to find an
exact  instanton solution. Therefore, in what follows we will restrict
ourselves to the case of weak dissipation, where a perturbative analysis can be
made. It is obvious that the opposite limit, the overdamped tunneling
``motion" of the Wigner crystal, is described by the trajectories with zero
winding number ($n=0$) and the Aharonov-Bohm oscillations vanish.

At first we consider the case of a weakly pinned Wigner crystal,
$\alpha V_{p} \ll T_{s}$, when the use of perturbation theory can be easily
justified. For a weakly pinned Wigner crystal which tunnels through the
impurity
homogeneously, it is convenient to treat it as an effective particle of mass
$M=Nm$ ($N$ is the total number of electrons in the ring) when calculating the
tunneling action. The characteristic time of tunneling, $\tau_{\ast}$, through
the barrier in this particular case is much smaller than the periods of time
between successive tunneling events, $\tau_{\ast}=a\sqrt{M/2V_{p}} \ll \beta$
[15] and since the effects of dissipation manifest itself most significantly on
large time scales ($\sim \beta$), one can reliably use perturbation theory.

Substituting unperturbed tunneling trajectories
$\varphi(\tau)=2 \pi \tau / \tau_{\ast},\;\; \tau\in [-\tau_{\ast}/2 ,
\tau_{\ast}/2]$ into Eqs.(1) and (15) one finds
\begin{equation}
A_{t}=\sqrt{2\pi\frac{V_{p}}{T_{0}}} \;+\; 0.3\ \frac{\hbar\gamma}{T_{0}} ,
\end{equation}
where the first term represents the tunneling action in the dissipationless
regime, $T_{0}\equiv \hbar v_{F}/L \;(V_{p} \agt T_{0})$, and the dissipative
correction $(\gamma\equiv\eta /2m, \; \hbar\gamma \ll T_{\ast} \simeq
\sqrt{V_{p} T_{0}})$ can be interpreted as being due to the broadening of the
quantized levels of the system.

In the dilute instanton gas approximation ($A_{t} \gg 1$) the amplitude of the
persistent current is proportional to $ \exp(- A_{t}) $. Hence dissipation
exponentially suppresses the Aharonov-Bohm oscillations (as it should, see
also [22, 12]). Notice that Eq.(16) is valid up to the crossover temperature
$T_{\ast} \simeq \sqrt{V_{p} T_{0}}$ (temperature corrections are of the order
$T/T_{\ast}$). Therefore the criterion for the perturbation
expansion to be valid, $\hbar\gamma \ll T_{\ast}$, does not have to be
violated in the case of moderately weak pinning ($V_{p} \agt T_{0} \rightarrow
T_{\ast} \agt T_{0}$) even if dissipation is strong (in terms of free
electrons) $\hbar\gamma\agt T_{0}$. The dissipation, of course,
remains weak for plasmons $\hbar\gamma\ll T_{s}$.

In the regime of strong pinning, $\alpha V_{p} \gg T_{s}$, the tunneling of a
Wigner crystal is inhomogenous and the characteristic tunneling time is of the
order of $L/s$. In this case one would expect a stronger influence of
dissipation on tunneling. However, it is easy to show that for a stiff ($\alpha
\ll 1$) crystal, the impact of dissipation on the persistent current is similar
to the case described above. By substituting Eq.(9) into Eq.(15) and doing the
integrals one can readily get the simple result
\begin{equation}
\Delta A_{t}
= \frac{1}{2}\ \frac{\hbar\gamma}{T_{0}} .
\end{equation}
Again the perturbative
correction due to dissipation does not depend on temperature or barrier height.
This expression exactly coincides with the dissipative action for an elementary
transition ($\Delta\varphi=\pm 2\pi, n=\pm 1$) of a perfect Wigner crystal
\begin{equation}
A_{t}=\frac{\pi}{2}\frac{T}{T_{0}} n^{2} +
\frac{1}{2}\frac{\hbar\gamma}{T_{0}}|n|
\end{equation}
This result can easily be obtained from Eqs.(1) and (15), by making use of the
extremal trajectories $\varphi_{n}=2\pi n\tau/ \beta$ for a perfect Wigner
crystal-ring. One can see from Eq.(18) that dissipative corrections are small
at
temperatures $T \gg \hbar\gamma$. Therefore in the high temperature region the
impact of dissipation on the persistent current --- even for a perfect Wigner
crystal --- can be accounted for perturbatively. The presence of a barrier
results in the appearence of a new energy scale and this enables us to use
perturbation theory even at $T \rightarrow 0$.

Thus, in all cases considered, the dissipative corrections to the persistent
current factorize as an overall factor, $ \exp(-const \;\hbar\gamma/T_{0}) $.
They  depend neither on temperature nor on the strength of the pinning
potential
(we assume that $ V_{p} \agt \hbar v_{F}/L$ ). This implies that dissipation,
even if the ring is strongly connected to reservoirs  ($\hbar\gamma \sim \hbar
v_{F}/L$), though it suppresses the current, does not change the anomalous
temperature behaviour predicted in Ref.~[15].

This work was supported by the Swedish Royal Academy of Sciences,
the Swedish Natural Science Research Council, and by grant U2K000
from the International Science Foundation.
One of us (I.K.) acknowledges the hospitality of the Department of Applied
Physics, CTH/GU.

\newpage

\begin{center}
{\bf REFERENCES}
\end{center}

\noindent [1] S.~Washburn, and R.~A.~Webb, {\em Repts. Progr. Phys.} {\bf 55},
 1311 (1992); \\
\noindent [2] V. Chandrasekhar, R.~A.~Webb, M.~J.~Brady, M.~B.~Ketchen,
W.~J.~Gallagher, and \\
 A.~Kleinsasser, {\em Phys. Rev. Lett.} {\bf 67}, 3578 (1991). \\
\noindent [3] D. Mailly, C.~Chapelier, and A.~Benoit, {\em Phys. Rev. Lett.}
{\bf 70}, 2020 (1993). \\
\noindent [4] D. Loss, {\em Phys. Rev. Lett.} {\bf 69}, 343 (1992). \\
\noindent [5] I. O. Kulik, {\em JETP Lett.} {\bf 11}, 275 (1970). \\
\noindent [6] H.-F. Cheung, Y. Gefen, E. K. Riedel, and W.-H. Shih,
{\em Phys. Rev.} \\
{\bf B37}, 6050 (1988). \\
\noindent [7] A. M\"{u}ller-Groeling, H. A. Weidenm\"{u}ller, and
C. H. Lewenkopf, \\
{\em Europhys. Lett.} {\bf 22}, 193 (1993). \\
\noindent [8] B.~S.~Shastry, and B.~Sutherland, {\em Phys. Rev. Lett.}
{\bf 65}, 243 (1990). \\
\noindent [9] C.~A.~Stafford, and A.~J.~Millis, {\em Phys. Rev.} {\bf B48},
1409 (1993).\\
\noindent [10] M.~Abraham, and R.~Berkovits, {\em Phys. Rev. Lett.}
{\bf 70}, 1509 (1993).\\
\noindent [11] I.~O.~Kulik, A.~S.~Rozhavsky, and E.~N.~Bogachek,
{\em JETP Lett.} {\bf 47}, 303 (1988). \\
E. N. Bogachek, I.~V.~Krive, I.~O.~Kulik,and A.~S.~Rozhavsky, \\
{\em Phys. Rev.} {\bf B42},7614 (1990). \\
\noindent [12] I. V. Krive, and A. S. Rozhavsky, {\em Int. J. Mod. Phys.}
{\bf B6}, 1255 (1992). \\
\noindent [13] D. V. Averin, and K. K. Likharev, In:
{\em Mesoscopic Phenomena
in Solids}, \\
 Eds. B. Altshuler, P. A. Lee, and R. A. Webb ( Elsevier,
Amsterdam, 1991). \\
K. Jauregui, W. H\"{a}usler, and B. Kramer, {\em Europhys. Lett.}
{\bf 24}, 581 (1993). \\
\noindent [14] L. I. Glazman, I. M. Ruzin, and B.~I.~Shlovskii,
{\em Phys. Rev.} {\bf B45}, 8454 (1992). \\
\noindent [15] I.~V.~Krive, R.~I.~Shekhter, S.~M.~Girvin, and M.~Jonson,
{\em G{\"o}teborg preprint} {\bf APR 93-38} (cond-mat/9404072);
I.~V.~Krive, R.~I.~Shekhter, S.~M.~Girvin, and M.~Jonson,
{\em G{\"o}teborg preprint} {\bf APR 93-53} \\
\noindent [16] A.~Furusaki, and N.Nagaosa,
{\em Phys. Rev.} {\bf B47}, 3827 (1993). \\
\noindent [17] H.~J.~Schulz, {\em Phys. Rev. Lett.} {\bf 71}, 1864 (1993); \\
\noindent [18] M.~Fabrizio, A.~O.~Gogolin, and S.~Scheidl,
{\em Phys. Rev. Lett.} {\bf 72},
2235 (1994). \\
\noindent [19] A. I. Larkin, and P. A. Lee, {\em Phys. Rev.} {\bf B17},
1596 (1978). \\
\noindent [20] A.~O.~Caldeira, and A.~J.~Leggett, {\em Ann. Phys.}
{\bf 149}, 374 (1983).\\
\noindent [21] G.~Sch\"{o}n, and A.~D.~Zaikin, {\em Phys. Repts.} {\bf 198},
237 (1990). \\
\noindent [22] M.~B\"{u}ttiker, {\em Phys. Rev.} {\bf B32}, 1846 (1985).

\newpage

\begin{center}
{\bf FIGURE CAPTION}
\end{center}

FIG.1. Temperature dependence of the normalized persistent current in a
strongly pinned Wigner crystal for two different stiffnesses, $\alpha^{-1}$,
without long-range Coulomb forces (solid line) and when Coulomb forces are
considered (dotted line). The Coulomb forces increase the effective stiffness
of
the Wigner lattice ($\alpha^{-1} \rightarrow \alpha^{-1}_{0} \sqrt{\ln N} $  ,
N
= the number of particles. Here N = 100) resulting in additional suppression of
the persistent current. The peak in the current will therefore become more
pronounced and the corresponding temperature will be shifted by a factor
$\simeq
\sqrt{\ln N}$. Note that the persistent current in all cases is normalized to
its zero temperature value, which is smaller in the presence of long-range
Coulomb forces (see text).

\end{document}